\documentclass[%
 reprint,
 amsmath,amssymb,
 aps,
]{revtex4-1}

\usepackage{graphicx}% Include figure files
\usepackage{dcolumn}% Align table columns on decimal point
\usepackage{bm}% bold math
\usepackage{color}

\begin{document}

\preprint{APS/123-QED}

\title{Prediction of Born effective charges using neural network to study ion migration under electric fields: applications to crystalline and amorphous Li$_3$PO$_4$}

\author{Koji Shimizu$^1$}
\email{shimizu@cello.t.u-tokyo.ac.jp}
\author{Ryuji Otsuka$^1$}
\author{Masahiro Hara$^1$}
\author{Emi Minamitani$^2$}
\author{Satoshi Watanabe$^1$}
\email{watanabe@cello.t.u-tokyo.ac.jp}

%\author{Takanori Moriya$^1$}

\affiliation{
$^1$Department of Materials Engineering, The University of Tokyo, 7-3-1 Hongo, Bunkyo-ku, Tokyo 113-8656, Japan \\
$^2$The Institute of Scientific and Industrial Research, Osaka University, 8-1 Mihogaoka, Ibaraki, Osaka, 567-0047, Japan
}

\date{ \today }

%%%%%%%%%
%アブストラクト
%%%%%%%%%
\begin{abstract}
Understanding ionic behaviour under external electric fields is crucial to develop electronic and energy-related devices using ion transport.
In this study, we propose a neural network (NN) model to predict the Born effective charges of ions along an axis parallel to an applied electric field from atomic structures.
The proposed NN model is applied to Li$_3$PO$_4$ as a prototype.
The prediction error of the constructed NN model is 0.0376 $e$/atom. 
In combination with an NN interatomic potential, molecular dynamics (MD) simulations are performed under a uniform electric field of 0.1 V/\AA, whereby an enhanced mean square displacement of Li along the electric field is obtained, which seems physically reasonable. 
In addition, the external forces along the direction perpendicular to the electric field, originating from the off-diagonal terms of the Born effective charges, are found to have a nonnegligible effect on Li migration. 
Finally, additional MD simulations are performed to examine the Li motion in an amorphous structure. 
The results reveal that Li migration occurs in various areas despite the absence of explicitly introduced defects, which may be attributed to the susceptibility of the Li ions in the local minima to the electric field. 
We expect that the proposed NN method can be applied to any ionic material, thereby leading to atomic-scale elucidation of ion behaviour under electric fields.
\end{abstract}

\maketitle

%%%%%%%%%%%%
\section{Introduction}
%%%%%%%%%%%%
Ion migration inside various devices, such as all-solid-state batteries and atomic switches  \cite{Terabe, Sugiyama, Nishio}, is achieved by applying external forces from applied electric fields. 
Numerous studies elaborating the stability of materials and the mobility of ions have been conducted using theoretical calculations because these aspects are directly related to device performance. 
To further advance our understanding of the operating mechanisms of such ion-conducting devices, atomic-scale analyses of ionic motion in device operating circumstances, that is, under electric fields, are crucial. 

Assuming a linear response, external forces arising from applied electric fields can be estimated simply by multiplying the electric field vector by the valence states of the ions. 
In electronic state calculations, such as density functional theory (DFT) calculations, the valence states are often evaluated, for instance, using Mulliken charges from the coefficients of atomic orbitals \cite{Mulliken} or Bader charges using charge density distributions \cite{Henkelman}. 
By contrast, the Born effective charges are defined from the induced polarisation in a periodic system by their atomic displacements (see Fig. 1(a)), or are equivalently defined from the induced atomic forces with respect to the applied electric fields. 
As our current interest lies in analysing ion motion under applied electric fields, and the latter definition precisely corresponds to the target situation, Born effective charges, rather than static valence states, are the suitable physical quantities to evaluate the external forces acting on the ions.
In addition, the Born effective charges can be quantified as the number of each atom without the arbitrariness of the decomposition of the total charges.
In most cases, these per-atom quantities are compatible with the computational processes of dynamic calculations using the methods described below.

Recently, atomistic simulations using interatomic potentials constructed via machine learning (ML) techniques have gained increasing attention. 
Representative methods include the high-dimensional neural network potential (NNP) \cite{Behler}, Gaussian approximation potential \cite{Barok}, moment tensor potential \cite{Shapeev}, and spectral neighbour analysis potential \cite{Thompson}. 
Numerous studies have demonstrated that ML potentials optimised using DFT calculation data can predict various physical quantities comparable to those of DFT calculations at low computational costs \cite{Artrith2, Lee, Shimizu2, Watanabe}. 
Notably, in their applications to solid electrolyte materials, the predicted ionic conductivities agree well with both the DFT and experimental results \cite{Li1, Marcolongo}.

To consider the application of these ML potentials for dynamic calculations under applied electric fields, predictive models of ion charges are necessary to evaluate the external forces, as stated earlier. 
Prior studies have proposed neural network (NN) models to predict the charges of ions \cite{Artrith1, Ko}; however, these models evaluate long-range electrostatic interactions or nonlocal charge transfer through the predicted charge states of ions. 
Therefore, in this study, we proposed an NN-based model to predict the Born effective charges for given structures.
In combination with the conventional NNP, the proposed NN model was applied to dynamic simulations to evaluate the external forces under a uniform electric field.
Herein, we employed Li$_3$PO$_4$ as the prototype material, which is commonly used in the research on all-solid-state Li batteries \cite{Haruta, Okuno, Shimizu1}.  
We verified our scheme of dynamic calculations based on the proposed NN model by evaluating ion behaviour under applied electric fields.

%%%%%%%%%%%%%
\section{Methodology}
%%%%%%%%%%%%%

%%%%%%%
\begin{figure}
\includegraphics[bb=0 0 4320 2548, width=0.45 \textwidth]{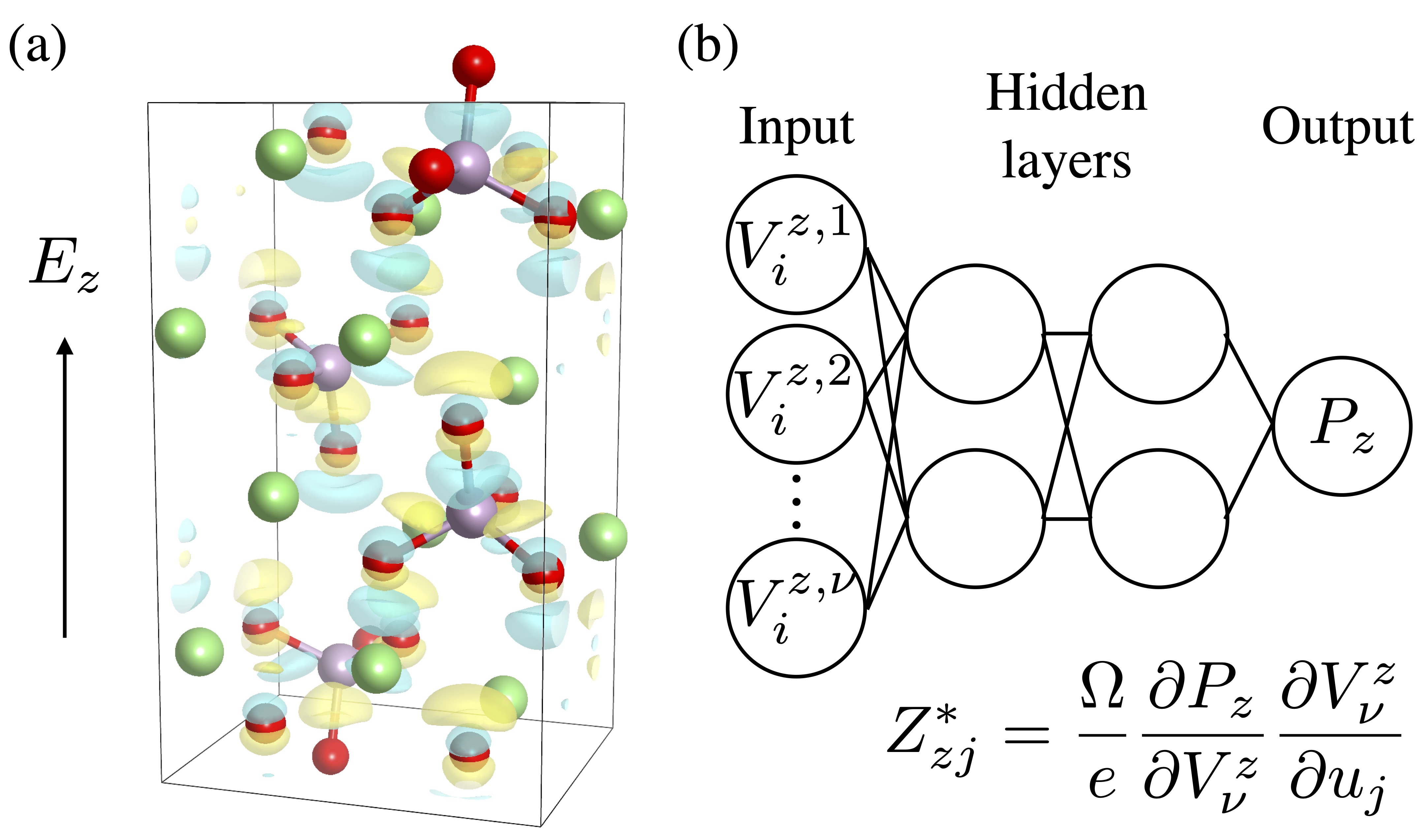}
\caption{(a) Schematic of electronic polarisation in Li$_3$PO$_4$ induced by an electric field along $z$-axis. The yellow (green) clouds depict the increased (decreased) parts of charge density differences, as visualised by VESTA software \cite{Momma}. (b) Schematic of the proposed NN model to predict Born effective charges.}
\label{fig1}
\end{figure}
%%%%%%%

Herein, we explain the computational details of the proposed NN model for the Born effective charge predictor. 
The Born effective charge is defined as follows.
\begin{equation}
Z^{*}_{ij} = \frac{\Omega}{e} \frac{\partial P_i}{\partial u_j} = \frac{1}{e} \frac{\partial F_i}{\partial E_j},
\label{eq1}
\end{equation}
where $\Omega$ and $e$ are the cell volume and the elementary charge, respectively;
$P_i$ and $u_j$ are the macroscopic polarisation and atomic coordinates, respectively;
$F_i$ and $E_j$ are the atomic forces and the electric field, respectively;
and subscripts $i$ and $j$ represent the $x$, $y$, or $z$ directions.
In the formalism of NNP, atomic forces can be obtained analytically by applying the chain rule in the following relation.
\begin{equation}
F_j = -\frac{\partial U}{\partial u_j} = -\frac{\partial U}{\partial G_\nu} \frac{\partial G_\nu}{\partial u_j},
\label{eq2}
\end{equation}
where $U$ is the total energy and $G_\nu$ represents the symmetry functions (SFs) \cite{Behler}. 
Based on the similarities in Eqs. (1) and (2), the framework of NNP appears to be modifiable as a Born effective charge predictor of a specific direction $i$ (one direction in the $3 \times 3$ tensor) by replacing $U$ and $F_j$ with $-\frac{\Omega}{e} P_i$ and $Z^{*}_{ij}$, respectively.
Essentially, we confirmed that the above modifications achieved some prediction performance; however, the obtained accuracy was not satisfactory. 
This inaccuracy can be rationalised by using the scalar quantities from the SFs as the inputs of NN to predict the vector quantities of macroscopic polarisation.

%%%%%%%
\begin{figure*}
\includegraphics[bb=0 0 5194 3496, width=0.8 \textwidth]{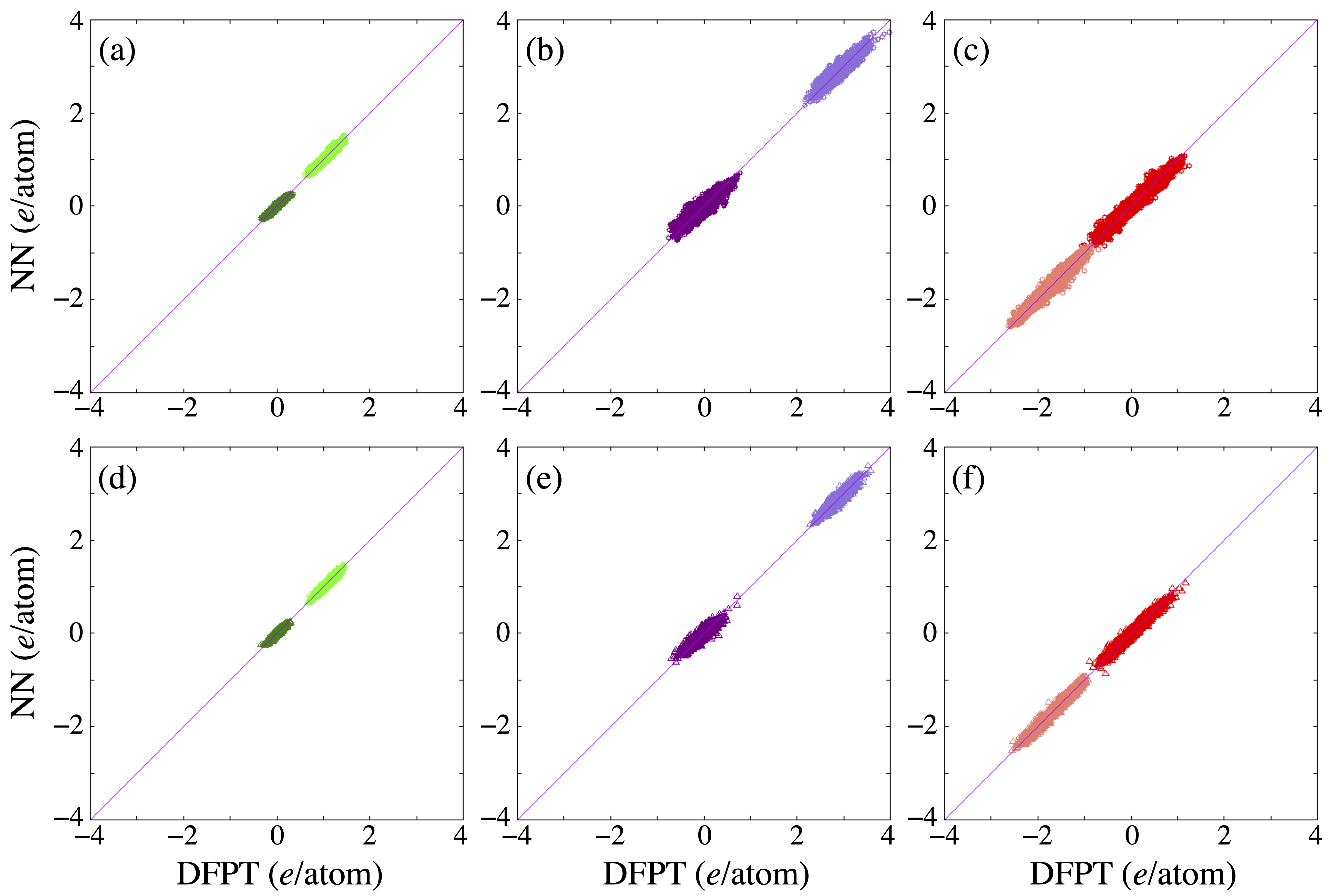}
\caption{Comparison between DFPT and NN on the Born effective charges of (a)-(c) training and (d)-(f) test sets. The comparisons are shown separately for each elemental species: (a, d) Li, (b, e) P, and (c, f) O. The light (dark) colours indicate the (off-)diagonal components.}
\label{fig2}
\end{figure*}
%%%%%%%

Hence, to preserve directional information in the inputs, we employed vector atomic fingerprint (VAF) \cite{Botu, Li2}, described below.
\begin{equation}
V^{1,\alpha}_i = \sum_j \frac{R^{\alpha}_{ij}}{R_{ij}} e^{-\eta (R_{ij} - R_s)} f_c(R_{ij}),
\label{eq3}
\end{equation}
\begin{equation}
\begin{split}
V^{2,\alpha}_i = & 2^{1-\zeta} \sum_j \sum_k ({\bf R}_{ij} + {\bf R}_{ik})^{\alpha} \{1+{\rm cos}(\theta_{ijk} - \theta_s) \}^{\zeta} \\
                        & e^{-\eta (\frac{R_{ij}+R_{ik}}{2} - R_s)^2} f_c(R_{ij}) f_c(R_{ik}),
\label{eq4}
\end{split}
\end{equation}
where $\eta$ and $\zeta$ are width parameters; 
$R_{ij}$ and $R_{ik}$ are the atomic vectors of atom $i$ with $j$ and $k$, respectively; 
$\theta_{ijk}$ is the angle between atoms $i$, $j$, and $k$ at vertex $i$;
$\alpha$ represents either the $x$, $y$, or $z$-coordinates;
$R_s$ and $\theta_s$ determine peak positions;
and $f_c$ is the cutoff function, which is expressed as
\begin{equation}
  f_c(R_{ij}) =
  \begin{cases}
    \frac{{\rm cos}(\frac{\pi R_{ij}}{R_c})+1}{2} & \text{if $R_{ij} \le R_c$,} \\
    0                                                     & \text{if $R_{ij} > R_c$,}
  \end{cases}
\label{eq5}
\end{equation}
where $R_c$ is the cutoff distance.
These functions are invariant to the rotation of the $\alpha$-axis. 
In a simplified model, the prediction of Born effective charge are sufficient only for a specific direction along the electric field (e.g., $zx$, $zy$, and $zz$ when $E_z$), as in the following expression (see Fig. 1(b)).
\begin{equation}
Z^{*}_{zj} = \frac{\Omega}{e} \frac{\partial P_z}{\partial V^{z}_{\nu}} \frac{\partial V^{z}_{\nu}}{\partial u_j}.
\label{eq6}
\end{equation}
Thus, the proposed NN model requires only a VAF with $\alpha = z$ as its input, which can be achieved by minimal modifications from the original NNP architecture.
Note that we may extend the model to predict Born effective charges in the form of tensors in future work.
Assuming that the forces vary linearly with the electric field, the external forces acting on the ions can be calculated as follows.
\begin{equation}
\Delta F_j^{\rm NN} = Z^{*}_{zj} E_z.
\label{eq7}
\end{equation}
The total forces were considered as the sum of the external forces and values obtained by the conventional NNP.
\begin{equation}
F^{\rm Total}_j = F^{\rm NNP}_j + \Delta F_j^{\rm NN}.
\label{eq8}
\end{equation}
Thus, simulations of ion dynamics under an electric field can be performed. 
The loss function of the proposed NN model includes errors in the macroscopic polarisation and Born effective charges in a manner similar to that of the NNP.
\begin{equation}
\begin{split}
\Gamma =~ & \alpha \sum_{n=1}^{N_{\rm Train}} \frac{ (P_{z, n}^{\rm NN} - P^{\rm DFPT}_{z, n})^2}{N_{\rm Train}} \\
                  & +~ \beta \sum_{n=1}^{N_{\rm Train}} \frac{ \Big\{ \sum_{m=1}^{N_i} \sum_{j \in \{ x,y,z \} } (Z^{* n, {\rm NN}}_{zj, m} - Z^{* n, {\rm DFPT}}_{zj, m} )^2 \Big\} }{3M_{\rm Train}},
\label{eq9}
\end{split}
\end{equation}
where $N_{\rm Train}$ and $M_{\rm Train}$ indicate the total amount of data and the total number of atoms, respectively. 
Here, training was executed by focusing on errors in the Born effective charges ($\alpha = 0$ and $\beta = 1$).
We performed atomistic simulations based on the proposed NN model using LAMMPS software \cite{Plimpton} with the homemade interfaces.

Next, we describe the training dataset of Li$_3$PO$_4$ used to construct the NNP model. 
Note that we used parts of the structures and the corresponding total energies and atomic forces of DFT calculations generated in Ref. \cite{Li1}. 
The dataset includes pristine (Li$_{12}$P$_4$O$_{16}$), Li vacancy (Li$_{11}$P$_4$O$_{16}$), and Li$_2$O vacancy (Li$_{22}$P$_8$O$_{31}$). 
For the pristine dataset, we used 14,001 snapshot structures of ab initio molecular dynamics (AIMD) with temperatures of 300, 2000, and 4000 K. 
The Li vacancy structures comprise 1,656 images from nudged elastic band (NEB) calculations. 
The Li$_2$O vacancy structures comprise 5,000 snapshots of AIMD at 2000 K. 
In total, 20,657 structures were used to construct NNP. 

As the training dataset for the proposed NN model, we used 8,000 Li$_{12}$P$_4$O$_{16}$ snapshots from AIMD calculations at temperatures of 300 and 2000 K, 5,000 Li$_{11}$P$_4$O$_{16}$ images from NEB calculations, and 4,900 Li$_{22}$P$_8$O$_{31}$ snapshots from AIMD calculations at 2000 K. 
For these 17,900 structures, we performed density functional perturbation theory (DFPT) calculations \cite{Gonze} to obtain the Born effective charge tensors. 
In the DFPT calculations, we used a generalized gradient approximation with the Perdew-Burke-Ernzerhof functional \cite{Perdew}, a plane-wave basis set of 500 eV cutoff energy, self-consistent field convergence criterion of $10^{-6}$ eV, and $k$-point sampling mesh of $6 \times 6 \times 4$ for Li$_{11}$P$_4$O$_{16}$ and $4 \times 4 \times 2$ for Li$_{12}$P$_4$O$_{16}$ and Li$_{22}$P$_8$O$_{31}$. 
All calculations were performed using the Vienna Ab initio Simulation Package \cite{VASP1, VASP2}. 
We used different structural datasets for the two models because the calculated Born effective charges of the strongly distorted structures resulted in unreasonably large values.
Additionally, as described above, the proposed NN model predicted three components (one direction of the $3 \times 3$ tensor) of the Born effective charge for simplicity. 
Thus, we trained the proposed NN model separately using the Born effective charges in each direction, considering the rotational manipulation of the structures and their tensor values, to enhance the variety of training datasets without performing additional DFPT calculations.

%%%%%%%%%%%%%%%%%%
\section{Results \& Discussion}
%%%%%%%%%%%%%%%%%%

%%%%%%%
\begin{figure}
\includegraphics[bb=0 0 2799 2676, width=0.5 \textwidth]{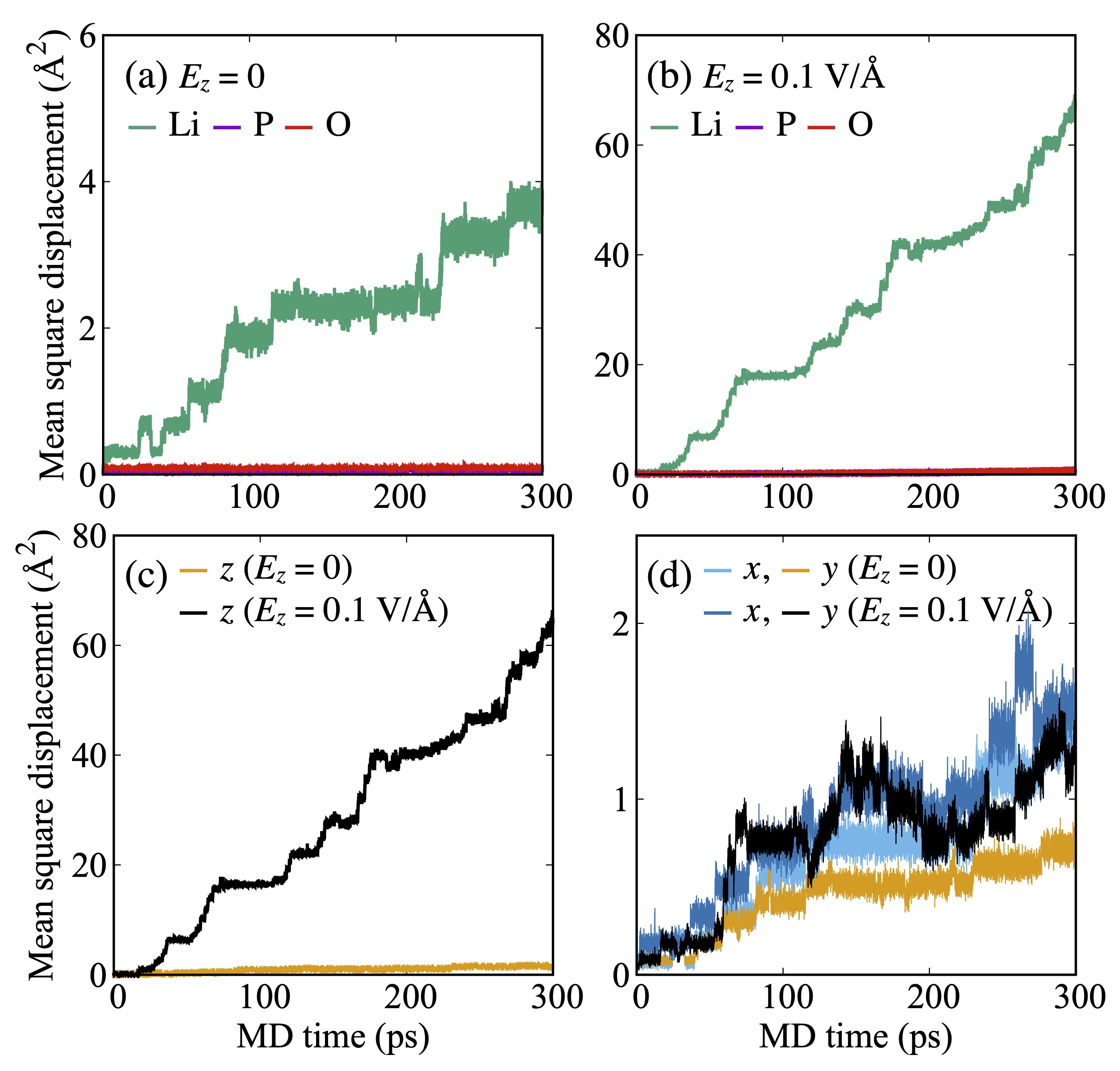}
\caption{Calculated MSDs of Li vacancy model (Li$_{47}$P$_{16}$O$_{64}$). The MD simulations with temperature of 800 K (a) without electric field and (b) with $E_z$ = 0.1 V/\AA, where the MSDs are separately shown for each element. The MSDs of Li are separately shown for (c) $x$ and $y$ and (d) $z$ components.}
\label{fig3}
\end{figure}
%%%%%%%

%%%%%%%
\begin{figure*}
\includegraphics[bb=0 0 4560 3136, width=0.8 \textwidth]{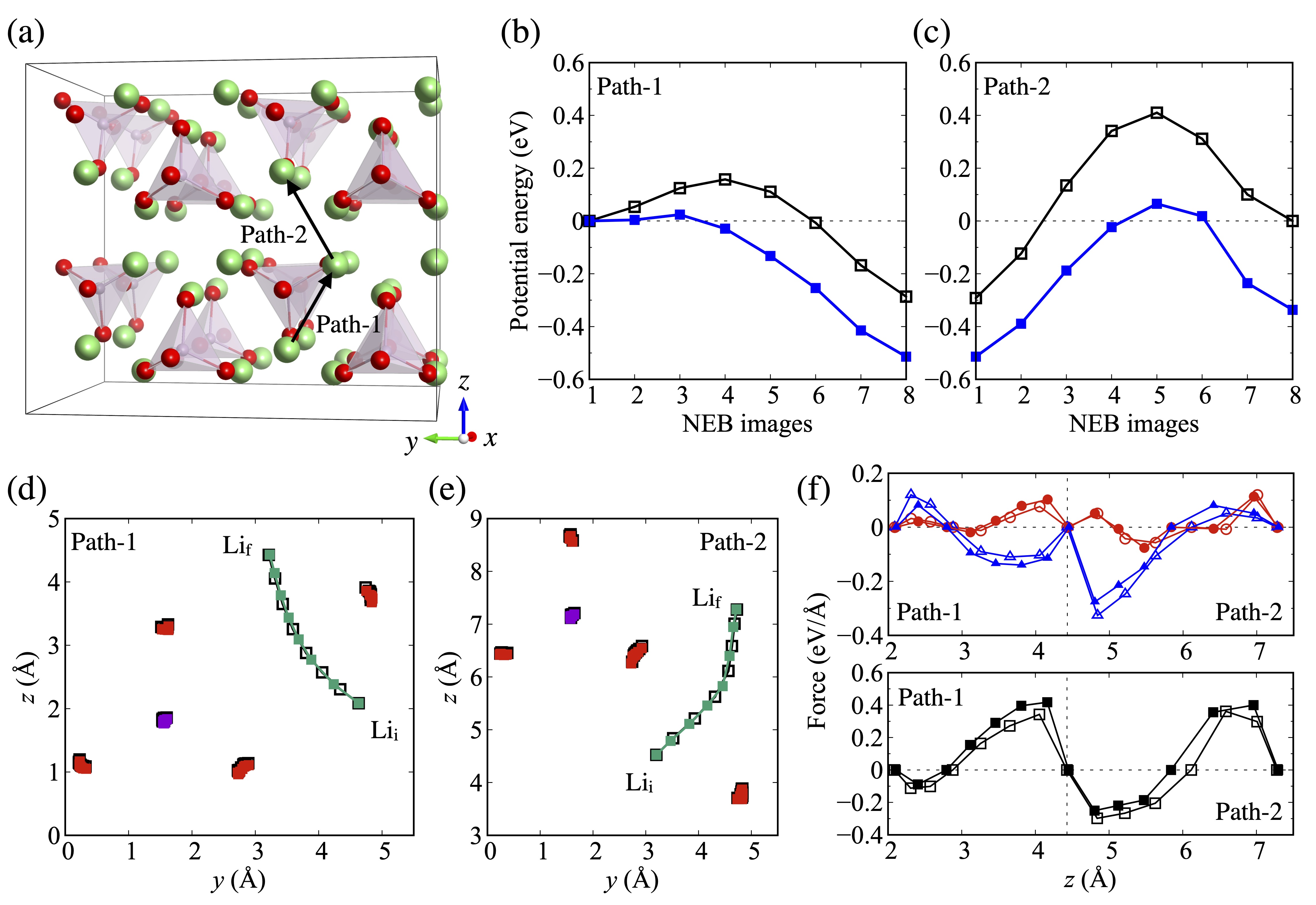}
\caption{NEB calculations of Li vacancy model (Li$_{47}$P$_{16}$O$_{64}$). (a) Schematic of Li vacancy migrating paths. The potential energy profiles with $E_z$ = 0 and $E_z$ = 0.1 V/\AA, respectively, for (b) Path-1 and (c) Path-2. The atomic coordinates of migrating Li and the neighbouring P and O atoms with $E_z$ = 0 and $E_z$ = 0.1 V/\AA, respectively, for (d) Path-1 and (e) Path-2. (f) Circles, triangles, and squares show the atomic forces of migrating Li for $x$-, $y$-, and $z$-directions, respectively, at each image as a function of its $z$-coordinate. The open and filled marks show $E_z$ = 0 and $E_z$ = 0.1 V/\AA, respectively.}
\label{fig4}
\end{figure*}
%%%%%%%

First, we constructed the NNP using a network architecture of 125 input nodes, two hidden layers with 15 nodes, and one output node, [125-15-15-1], for each elemental species. 
The root-mean-square errors (RMSEs) of the total energies and atomic forces were 3.34 (2.91) meV/atom and 86.1 (87.9) meV/\AA, respectively, for the randomly chosen 90\% (10\%) of the training (test) data. 
The obtained RMSE values were sufficiently small compared with those of other studies using NNP \cite{Li1, Marcolongo}. 
Please refer to Fig. S1 for a comparison between the NNP predictions and DFT reference values. 
The hyperparameters used in the SFs are listed in Tables S1 and S2.

Next, we constructed the proposed NN model for the Born effective charge predictor. 
We used the NN architecture of [180-10-10-1], where the RMSEs of the training (randomly chosen 90\%) and test (remaining 10\%) data were 0.0378 $e$/atom and 0.0376 $e$/atom, respectively. 
%Note that また、ひとまわり大きなセルについてもRMSEが同程度であることを確認した。
Tables S3 and S4 present the hyperparameters used in VAFs. 
Figure 2 compares the predicted Born effective charges and their DFPT values.
Evidently, all the points, including both the diagonal and off-diagonal components, were located near the diagonal lines, thus suggesting fairly good predictions. 
In addition, the distributions show that the diagonal components of the Born effective charges varied considerably from their formal charges, that is, Li: $+1$, P: $+3$, and O: $-2$, owing to the structural changes. 
Moreover, the charge states of oxygen underwent the largest variation, despite the fact that Li is a mobile species in Li$_3$PO$_4$. 
Furthermore, such variations can be observed in the off-diagonal components of the Born effective charges, although these values were typically small in the crystalline structure: the averages of ${ \sum_{i \ne j} \sqrt{ Z^{* 2}_{ij} }/6 }$ for Li, P, and O are $3.87 \times 10^{-2}$, $2.47 \times 10^{-3}$, and $1.59 \times 10^{-1}$, respectively.
In particular, the off-diagonal values of oxygen varied the most between $-1$ and $+1.5$.
In the following dynamic calculations, we used a uniform electric field of $E_z$ = 0.1 V/\AA~applied in the $z$ direction. 
Although the magnitude of the electric field may be excessively large compared with the actual device operating conditions, we chose that value to magnify its effect within a feasible computational time. 
In addition, the errors of the external forces at the magnitude of this electric field could be estimated to be on the order of meV/\AA, suggesting that the constructed NN model was sufficiently accurate.

Using both the constructed NNP and the proposed NN model, we performed canonical ensemble ($NVT$) MD simulations to investigate ion motion under an electric field. 
For these MD simulations, the temperature and computation time were set to 800 K and 300 ps, respectively. 
Based on the prior knowledge that Li moves through vacancy sites, we used a crystalline Li$_{47}$P$_{16}$O$_{64}$ model, which contained one Li vacancy (V$_{\rm Li}$) in the supercell. 
Note that the size of the simulation model was larger than that of the training dataset, that is, it had a lower V$_{\rm Li}$ concentration. 
In fact, Li ions seldomly moved in the MD simulations using the pristine model with the above settings or the Li vacancy model at temperatures lower than 800 K.

Figure 3(a) shows the mean square displacement (MSD) of each elemental species, calculated from the trajectories of the MD simulations without an electric field. 
The MSD of Li increased slightly with MD time, whereas those of P and O remained nearly zero, thus indicating that these elemental species were immobile. 
In the calculated MSDs under the electric field shown in Fig. 3(b), the MSD of Li increased rapidly, thus indicating Li migration. 
By contrast, the MSDs of P and O remained small ($< 1~{\rm \AA}^2$), as in the case without an electric field. 
Figures 3(c) and (d) show the MSDs of Li in each direction. 
We attribute the rapid growth of the total MSD under the electric field to the contributions from the $z$-direction, which is a physically reasonable result, as it is consistent with the direction of the electric field. 
In addition, the Li migration paths in crystalline Li$_3$PO$_4$ are not always straight along the $z$ direction, considering that Li moves along the lattice sites via the vacancy hopping mechanism. 
Hence, the MSDs along the $x$- and $y$-directions under an electric field exhibited more fluctuating behaviour compared with the case of without an electric field.

We performed NEB calculations to examine the changes in the potential energy profiles in the presence of electric field. 
Here, the electrostatic energy that the moving V$_{\rm Li}$ acquires from the electric field, $Z^{*}_{zz} E_z \Delta z$, is added to the total energy.
For all the NEB calculations, we set the number of intermediate images to six.
Figure 4(a) shows two migration paths of Li, in which Li moved primarily along the $z$-direction. 
The potential energy profiles obtained for the two paths are shown in Figs. 4(b) and (c). 
For both paths, we set the potential energy of the initial structure in Path-1 as the reference. 
In the case of Path-1, the potential energy barrier of Li migration (from images 1 to 8) was reduced from 0.157 to 0.0242 eV by the electric field. 
By contrast, the potential energy barrier in the opposite direction (images 8 to 1) increased from 0.444 to 0.485 eV. 
Similarly in the case of Path-2, the potential energy barrier of Li migration from images 1 to 8 (8 to 1) decreased from 0.702 (0.410) to 0.579 (0.402) eV by the presence of electric field. 
This directionality of the electric field affects the potential energies and facilitates Li migration along the direction of the electric field. 
In addition, we confirmed that this directionality was almost invisible along a path nearly perpendicular to the electric field (please refer to Fig. S2). 
The observed decrease in the potential energy barrier and directionality of ion migration agreed with previous NEB calculations of O defects in MgO using the modern theory of polarisation \cite{El-Sayed}.

The atomic coordinates of the migrating Li and the surrounding P and O atoms with or without the electric field in Paths-1 and -2 are shown in Figs. 4(d) and (e), respectively. 
Note that for comparison, we set the initial atomic positions of the migrating Li in the two cases to be identical. 
We found that the intervals between intermediate images varied according to the presence of an electric field, whereas the paths were similar. 
Figure 4(f) shows the atomic forces acting on the migrating Li in each NEB image as a function of the $z$ coordinates. 
Evidently, the absolute values of the atomic forces along the $z$-direction decreased and increased in front of and behind the potential energy barrier, respectively. 
This indicates a decrease in the barrier height and, subsequently, an acceleration of the Li motion along the $z$ direction. 
Moreover, we found that the atomic forces along $y$-direction shifted negatively and positively in the presence of an electric field over Paths-1 and -2, respectively. 
Because these shifts correspond to the direction of Li motion along $y$-axis, we considered that the external forces facilitated the movement of Li toward the final positions, whereas their contribution was not as large as those in the $z$ direction. 
By contrast, the changes in the atomic forces along the $x$ direction were minor because the migrating Li in these paths moved primarily along the $z$ direction, followed by $y$ direction.

Subsequently, we performed additional MD simulations to further examine the effects of external forces along $x$- and $y$-directions on Li migration, that is, the contribution from the off-diagonal components of the Born effective charges with respect to the electric field in the $z$-direction. 
In these simulations, we considered only the $Z_{zz}^*$, and the $Z_{zx}^*$ and $Z_{zy}^*$ were set to 0 to exclude their contributions. 
Thus, we obtained a considerably smaller MSD for Li than that shown in Fig. 3(b) (see Fig. S3). 
A comparison of the MD trajectory lines shown in Fig. S4 indicates pronounced Li migratory behaviour when all three terms are considered. 
These results suggest that the off-diagonal components slightly but effectively confined the potential energy surface of the Li migration paths and consequently enhanced Li motion.
This enhancement did not appear when the charges were treated as scalar quantities because of the absence of external forces along the $x$- and $y$-directions.
This also demonstrates the significance of using the Born effective charges to study ion behaviour under electric fields.

%%%%%%%
\begin{figure}
\includegraphics[bb=0 0 3844 5688, width=0.5 \textwidth]{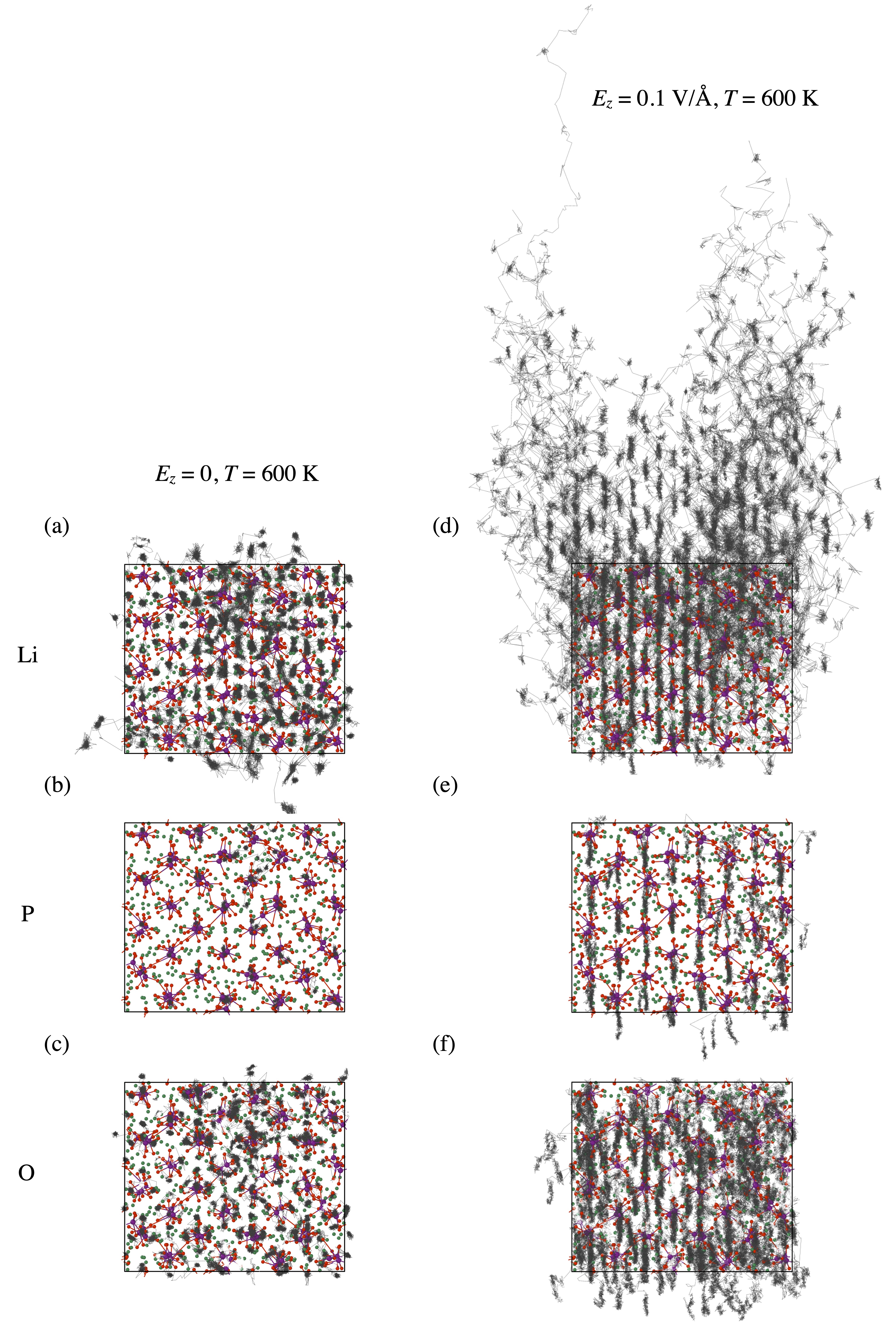}
\caption{Calculated MD trajectory lines of (a, d) Li, (b, e) P, and (c, f) O in the amorphous Li$_3$PO$_4$ model. The upward direction in the figure indicates the $z$-direction. The MD simulations are performed for 300 ps with the temperature of 600 K (a)-(c) without and (d)-(f) with the electric field.}
\label{fig5}
\end{figure}
%%%%%%%

Finally, we performed MD simulations with amorphous Li$_3$PO$_4$ under the electric field using the proposed scheme. 
An amorphous structure was generated using the melt-quench approach, as described in Ref. \cite{Li1}, and the model contained 384 Li, 128 P, and 512 O atoms without including specific defects. 
Please refer to Fig. S5(a) for the structural image. 
Here, we set the temperature to 600 K, which is lower than that used in the above cases. 
Without an electric field, the ions were displaced only to the optimal positions at the early stage of MD, as indicated by the MSDs shown in Fig. S5(b). 
By contrast, we observed considerably high mobility of Li under an electric field. 
The obtained MSD value shown in Fig. S5(c) is higher than that shown in Fig. 3(b). 
Figure 5 shows the trajectory lines of each ion in these two cases. 
The results clearly show that the ions were immobile without an electric field, whereas the Li ions moved extensively along the $z$ direction in the presence of an electric field. 
Notably, the P and O ions became relatively mobile in the amorphous structure compared with crystalline Li$_3$PO$_4$. 
Furthermore, we found that Li ions moved across the entire region in the amorphous model, whereas Li hopping was restricted to the vacancy sites in the crystal. 
We also noted that the Li ions moved more pronouncedly at 800 K, as shown in Fig. S6.
These results suggest that the Li ions located at the local minima of the metastable amorphous structure susceptibly undergo the electric field effects and readily overcome the mobility barrier.

%%%%%%%%%%%%
\section{Conclusion}
%%%%%%%%%%%%
We proposed an NN model to predict the Born effective charges of ions based on their atomic structures. 
We demonstrated the performance of our proposed NN model using Li$_3$PO$_4$ as a prototype ion-conducting material, where the error in the constructed model reached 0.0376 $e$/atom. 
In combination with the conventional NN potential, MD simulations were performed under a uniformly applied electric field. 
The obtained results indicated an enhanced displacement of Li along the electric field, which is physically reasonable. 
In addition, we confirmed the lowering of the potential energy barriers from NEB calculations under an electric field. 
Furthermore, we found that the external forces arising from the off-diagonal terms of the Born effective charges slightly but effectively confined the potential energy surface of the Li migration paths and consequently enhanced Li motion. 
Finally, we examined the Li behaviour in the amorphous Li$_3$PO$_4$ structure. 
We found that Li ions located at the local minima were susceptible to electric field effects and readily overcame the mobility barrier. 
These results suggest that the Born effective charge tensors, depending on the local atomic structures, may be a suitable quantity for a detailed analysis of ion behaviour under external electric fields.

%%%%%%%%%%%
Acknowledgements.
%%%%%%%%%%%
We thank Mr. Takanori Moriya for his contributions in the early stage of this study, and Editage (www.editage.com) for English language editing.
This study was supported by JST CREST Programs "Novel electronic devices based on nanospaces near interfaces" and "Strong field nanodynamics at grain boundaries and interfaces in ceramics" and JSPS KAKENHI Grant Numbers 19H02544, 20K15013, 21H05552, 22H04607, 23H04100. Some of the calculations used in this study were performed using the computer facilities at ISSP Supercomputer Center and Information Technology Center, The University of Tokyo, and Institute for Materials Research, Tohoku University.

%\end{acknowledgment}

\bibliography{apssamp}% Produces the bibliography via BibTeX.

\begin{thebibliography}{99}
\def\cms{Comput.\ Mater.\ Sci.\ }
\def\cp{Chem.\ Phys.\ }
\def\nat{Nature }
\def\pr{Phys.\ Rev.\ }
\def\pra{Phys.\ Rev.\ A }
\def\prb{Phys.\ Rev.\ B }
\def\prl{Phys.\ Rev.\ Lett.\ }
\def\jcp{J.\ Chem.\ Phys.\ }
\def\jpcm{J.\ Phys.: Condens.\ Matter }
\def\jpsj{J.\ Phys.\ Soc.\ Jpn.\  }
\def\susc{Surf.\ Sci.\ }
\def\suscrep{Surf.\ Sci.\  Rep.\ }
\def\pss{Prog.\ Surf.\ Sci.\ }
\def\ptps{Prog.\ Theor.\ Phys.\ Suppl.\ }
\def\zeitphysb{Z.\ Phys.\ B}



\bibitem{Terabe}
K. Terabe, T. Hasegawa, T. Nakayama and M. Aono, Nature {\bf 433}, 47 (2005).

\bibitem{Sugiyama}
I. Sugiyama, R. Shimizu, T. Suzuki, K. Yamamoto, H. Kawasoko, S. Shiraki and T. Hitosugi, APL Mater. {\bf 5}, 046105 (2017).

\bibitem{Nishio}
K. Nishio, T. Shirasawa, K. Shimizu, N. Nakamura, S. Watanabe, R. Shimizu and T. Hitosugi, ACS Appl. Mater. Interfaces {\bf 13}, 15746 (2021).

\bibitem{Mulliken}
R. Mulliken, J. Chem. Phys. {\bf 23}, 1833 (1955).

\bibitem{Henkelman}
G. Henkelman, A. Arnaldsson and H. J\'{o}nsson, Comput. Mater. Sci. {\bf 36}, 254 (2006).

\bibitem{Behler}
J. Behler and M. Parrinnelo, Phys. Rev. Lett. {\bf 98}, 146401 (2007).

\bibitem{Barok}
A. Bar\'{o}k, M. Payne, R. Kondor and G. Csányi, Phys. Rev. Lett. {\bf 104}, 136403 (2010).

\bibitem{Shapeev}
A. Shapeev, Multiscale Model. Simul. {\bf 14}, 1153 (2016).

\bibitem{Thompson}
A. Thompson, L. Swiler, C. Trott, S. Foiles and G. Tucker, J. Compt. Phys. {\bf 285}, 316 (2015).

\bibitem{Artrith2}
N. Artrith, J. Phys. Energy {\bf 1}, 032002 (2019).

\bibitem{Lee}
D. Lee, K. Lee, D. Yoo, W. Jeong and S. Han, Comput. Mater. Sci. {\bf 181}, 109725 (2020).

\bibitem{Shimizu2}
K. Shimizu, E. Arguelles, W. Li, Y. Ando, E. Minamitani and S. Watanabe, Phys. Rev. B {\bf 103}, 094112 (2021).

\bibitem{Watanabe}
S. Watanabe, W. Li, W. Jeong, D. Lee, K. Shimizu, E. Minamitani, Y. Ando and S. Han, J. Phys. Energy {\bf 3}, 012003 (2021).

\bibitem{Li1}
W. Li, Y. Ando, E. Minamitani and S. Watanabe, J. Chem. Phys. {\bf 147}, 214106 (2017).

\bibitem{Marcolongo}
A. Marcolongo, T. Binninger, F. Zipoli and T. Laino, Chem. Syst. Chem. {\bf 2}, e1900031 (2019).

\bibitem{Artrith1}
N. Artrith, T. Morawietz and J. Behler, Phys. Rev. B {\bf 83}, 153101 (2011).

\bibitem{Ko}
T. Ko, J. Finkler, S. Goedecker and J. Behler, Nat. Commun. {\bf 12}, 398 (2021).

\bibitem{Haruta}
M. Haruta, S. Shiraki, T. Suzuki, A. Kumatani, T. Ohsawa, Y. Takagi, R. Shimizu and T. Hitosugi, Nano Lett. {\bf 15}, 1498 (2015).

\bibitem{Okuno}
Y. Okuno, J. Haruyama and Y. Tateyama, ACS Appl. Energy Mater. {\bf 3}, 11061 (2020).

\bibitem{Shimizu1}
K. Shimizu, W. Liu, W. Li, S. Kasamatsu, Y. Ando, E. Minamitani and S. Watanabe, Phys. Rev. Mater. {\bf 4}, 015402 (2020).

\bibitem{Botu}
V. Botu and R. Ramprasad, J. Quantum Chem. {\bf 115}, 1074 (2015).

\bibitem{Li2}
W. Li and Y. Ando, Phys. Chem. Chem. Phys. {\bf 20}, 30006 (2018).

\bibitem{Plimpton}
S. Plimpton, J. Comp. Phys. {\bf 117}, 1 (1995).

\bibitem{Gonze}
X. Gonze and C. Lee, Phys. Rev. B {\bf 55}, 10355 (1997).

\bibitem{Perdew}
J. Perdew, K. Burke and M. Ernzerhof, Phys. Rev. Lett. {\bf 77}, 3865 (1996).

\bibitem{VASP1}
G. Kresse and J. Furthm\"{u}ller, Comput. Mater. Sci. {\bf 6}, 15 (1996).
\bibitem{VASP2}
G. Kresse and J. Furthm\"{u}ller, Phys. Rev. B {\bf 54}, 11169 (1996).

\bibitem{El-Sayed}
A.-M. El-Sayed, M. Watkins, T. Grasser and A. Shluger, Phys. Rev. B {\bf 98}, 064102 (2018).

\bibitem{Momma}
K. Momma and F. Izumi, J. Appl. Cryst. {\bf 44}, 1272 (2011).


\end{thebibliography}

\end{document}